\documentclass[12pt]{article}
\usepackage[pdftex]{graphicx}
\usepackage[cmex10]{amsmath}
\usepackage[margin=2cm]{geometry}
\usepackage{latexsym}
\usepackage{algorithm, algorithmic} %, algorithmic-fix}
\usepackage{amsfonts}
\usepackage[subrefformat=parens,labelformat=parens]{subfig}
\usepackage{setspace}

\usepackage{amsmath}%,}
\long\def\comment #1\commentend{}
\begin{document}

\newcommand{\bb}[1]{\mathbf{#1}}
\newcommand{\ba}[1]{\mathsf{#1}}
\newcommand{\ca}[1]{{\boxed{\mathsf{#1}}}}
\newcommand{\recipe}[1]{\ca{#1} \longrightarrow}
\newcommand{\subactions}[1]{\mathsf{#1}}
\date{}

\title{%A Statistical and Data Mining Analysis of WhatsApp
A Study of WhatsApp Usage Patterns and Prediction Models without Message Content\thanks{This research is based on work supported in part by MAFAT
and the ISRAEL SCIENCE FOUNDATION grant \#1488/14.}}

\author{Avi Rosenfeld$^{1}$, Sigal Sina$^{2}$, David Sarne$^{2}$\\ Or Avidov$^{2}$, Sarit Kraus$^{2}$ \\
       $^{1}$Department of Information Systems\\ Jerusalem College of Technology, Israel \\
       $^{2}$Department of Computer Science\\ Bar-Ilan University, Israel\\
       rosenfa@jct.ac.il, sigal.sina@gmail.com, sarne@cs.biu.ac.il\\ oravi90@gmail.com, sarit@cs.biu.ac.il}

\maketitle

\begin{abstract}
Internet social networks have become a ubiquitous application allowing people to easily share text, pictures, and audio and video files. Popular networks include WhatsApp, Facebook, Reddit and LinkedIn. We present an extensive study of the usage of the WhatsApp social network, an Internet messaging application that is quickly replacing SMS messaging. In order to better understand people's use of the network, we provide an analysis of over 6 million messages from over 100 users, with the objective of building demographic prediction models using activity data. We performed extensive statistical and numerical analysis of the data and found significant differences in WhatsApp usage across people of different genders and ages.  We also inputted the data into the Weka data mining package and studied models created from decision tree and Bayesian network algorithms. We found that different genders and age demographics had significantly different usage habits in almost all message and group attributes.  We also noted differences in users' group behavior and created prediction models, including the likelihood a given group would have relatively more file attachments, if a group would contain a larger number of participants, a higher frequency of activity, quicker response times and shorter messages. We were successful in quantifying and predicting a user's gender and age demographic. Similarly, we were able to predict different types of group usage. All models were built without analyzing message content. We present a detailed discussion about the specific attributes that were contained in all predictive models and suggest possible applications based on these results.

\end{abstract}
\doublespacing
\section{Introduction}
Internet social networks have become a ubiquitous application allowing people to easily share text, pictures, and audio and video files. Popular networks include Facebook, Reddit and LinkedIn, all of which maintain websites which serve as hubs facilitating people's information sharing. In contrast, the relatively new WhatsApp application is a smartphone application that enables people to share information directly via their phones. Since its introduction in 2009, its growth has steadily increased, and as of April 2016, it numbers over a billion monthly active users.\footnote{http://www.wired.com/2016/02/one-billion-people-now-use-whatsapp/} While many alternatives to WhatsApp are currently available in different online application stores (e.g., Kik, Telegram, Line Messenger, BBM, WeChat), WhatsApp is currently the most popular messaging application with the largest name recognition, by far the largest user base, and the strongest corporate backing since its acquisition by Facebook in 2014.  Given the emerging importance of this network it is not surprising that there is a growing interest in researching it, including user studies about people's WhatsApp use and possible applications \cite{Jain2016,Gulacti2016,Fiaa,Church2013,Pielot2014,OHara2014,Buchnik2014,Mudliar2015,Montag2015,Johnston2015}.

This paper's main contribution is that we have successfully created models that predict usage patterns between different types of users and groups without relying on the content of people's text messages. Prior WhatsApp work, as discussed in more detail in the following section, typically based its analysis on the content within the messages \cite{Wang2013,Argamon2009,WagnerGJS15}. Collecting and storing text messages is problematic for several reasons. First, privacy concerns exist in storing and analyzing people's messages and can raise significant ethical concerns \cite{van2004ethical}. Second, storing all information within peoples' text can require large amounts of storage, which in turn increases the cost of such analyses \cite{fan2006tapping}.  Instead, we exclusively focus on general message information such as the message's length, the size of the conversation group to which it was sent and temporal properties such as the time it was sent and how much time elapsed between this message and the previous one.    Despite the lack of content, we successfully created models that predict usage patterns between different types of users and groups. In previous studies, such patterns were found by checking a specific thesis via distributing and analyzing targeted questions within questionnaires \cite{Church2013,Pielot2014,OHara2014,Mudliar2015}, something that is significantly more time-intensive than the automated machine learning approach that we used. While this methodology has been used to study other social networks, including Facebook \cite{Wang2013,Xiang2010,Bakshy2012} and MySpace \cite{Thelwall2010}, applying these tools to the WhatsApp network is significantly more complicated because no public dataset currently exists, in contrast to these other networks. This is likely because of the medium involved -- while other social networks are primarily web-based and thus given to compiling data through web crawling, the WhatsApp network is based on individuals' private phone use and thus not publicly available. Furthermore, these studies typically use the messages' actual content, something we intentionally did not use.

As we further describe in the following sections, we performed an in-depth study based on WhatsApp messages and conversation groups by collecting over 6 million WhatsApp messages from 111 students between the ages of 18 and 34. Our analysis of this data revealed several key insights. First, we did in fact find significant differences in WhatsApp usage across people of different genders and ages.   Second, we inputted the data into the Weka data mining package \cite{Weka} and studied the output from decision tree and Bayesian network algorithms. This was mainly as a proof of concept for the kind of results one may extract by applying machine learning and data mining tools on WhatsApp data when collected in the message level, without getting exposed to the content itself. Despite our lack of relying whatsoever on any user generated content, these algorithms were successful in building models that can accurately predict a person's gender and approximate age. They were also successful in predicting which WhatsApp groups have certain qualities, such as higher percentages of file attachments, quicker responses, larger discussion groups and shorter messages. One key advantage in analyzing the results from the decision tree algorithm is that it outputs an unbiased assessment about which attributes and logical rules were important in building these prediction models, thereby providing additional insights.  Last, we note the importance of these results with possible future directions and applications.

\section{Related Work}
The WhatsApp social network is unique in several ways relative to other social networks. This application was developed to allow users to privately and freely send messages to each other through their smartphones. It provides a free alternative to SMS (Short Message Services) which is often still a metered (pay per use) service. Not only is WhatsApp often more cost effective than SMS, but it facilitates large group conversations, something that is difficult through SMS, if not impossible. While freely sharing information over the Internet is common to many social networks, and other public messaging services, such as Twitter, exist, the private nature of the WhatsApp network makes it rather unique. A similar difference between WhatsApp and other social networks is that membership is created and updated directly via people's smartphones. Not only is registration done exclusively through one's phone number, but the smartphone is the primary interface for sending and receiving messages.\footnote{While we note that a computer interface for WhatsApp exists, it is exclusively an interface for people's smartphones and offers no additional functionality.} Third, WhatsApp interpersonal conversation groups are the network's only communication medium and are formed by adding people's telephone numbers to that group. In contrast, other social networks are based on user membership and primarily focus on public messages where these messages are sent to all connected users (i.e these messages are called \emph{Posts} on Facebook and \emph{Tweets} on Twitter), and not through private groups.  Furthermore, Facebook is a network for publicly sharing photos, updates, and general news with members who ``follow" you. Twitter is a microblog network where members interact through concise messages of up to 140 characters. Given these and other differences between WhatsApp and other social networks, we believe that existing research about other networks is not necessarily applicable, and a new and thorough analysis of WhatsApp is warranted.

Much recent work has been dedicated to the study of how people use WhatsApp and the role of this new application in social communication. Most works to date have analyzed peoples' behavior through conducting surveys and targeted interviews. For example, work by Church and Oliveira conducted an online study asking targeting questions of users which were aimed at understanding differences between WhatsApp and SMS usage \cite{Church2013}. Pielot et. al \cite{Pielot2014} created a survey focusing on the question whether people expected an answer to their WhatsApp and SMS messages within several minutes.  O'Hara et. al interviewed 20 WhatsApp users for nearly an hour each, asking them semi-structured questions aimed at determining the nature of relationships forged with the people with whom they communicated \cite{OHara2014}. Mudliar and Rangaswamy \cite{Mudliar2015} spent over 350 hours observing 109 students, as well as conducted surveys to understand gender differences within Indian students' use of WhatsApp. All of these  studies can be characterized as being formed through a desire to answer specific questions by conducting targeted surveys and interviews.

This work is unique in that it uses statistical and data mining methods to study WhatsApp usage at the message level even without knowing the content of these messages. Our study contains the same motivation of previous WhatsApp research in that we also analyze differences between genders, the time that elapses until a message is answered, and the characteristics of larger and smaller discussion groups. However, our study is fundamentally different in that we are based solely on actual WhatsApp meta-message data, in order to perform our analysis without any possible human bias. The issue of human bias within smartphone usage analysis was recently studied, and one of the study's conclusions was that people poorly  report  their own usage in questionnaires \cite{question2015}. To our knowledge, only one other study, performed by Montag et. al \cite{Montag2015}, studied WhatsApp usage through logging data from nearly 2500 participants. While the number of participants in this study is impressive, the actual data logged was significantly less robust than in this study as they only collected general meta-data about use, only limited information about WhatsApp messages and no information about users' group activity.

In theory, even more accurate models could have been constructed had we also analyzed the messages' content. Specifically, models have been previously developed which can  predict an author's gender, age, native language or personality \cite{Wang2013,Argamon2009} based on content.  Examples include work by Argamon \cite{Argamon2009} which focused on creating models that identify word usage differences between men and women on Internet blogs. Similarly, Wagner et. al \cite{WagnerGJS15} focused on content differences between men and women in Wikipedia, and Wang, Burke and Kraut performed a study of content differences between genders on Facebook \cite{Wang2013}. However, as the WhatsApp network is inherently private, such approaches could not be applied in our case due to privacy concerns. As we now detail, even despite not having this information we were indeed similarly successful in predicting a user's demographic and group behavior.

\section{Dataset Creation and Description}

Given the private nature of the WhatsApp network, this study's first challenge was to create a WhatsApp message dataset while still insuring users' privacy. To do so, we developed software that integrated with the Android Debug Bridge (the ADB is an external tool which is able to backup an Android application).\footnote{Both the ADB software developed and the data collected are available from the authors.} This enabled taking a ``snapshot" of a person's groups and messages as they appear in her phone.
In order to make the data anonymous, the software encrypts the data that was pulled directly from the participant's smartphone by using the HMAC hash function. The entire process of obtaining a participant's data lasted approximately 15 minutes and we compensated each participant \$12 for their time and temporary inability to use their phones. We also collected the participants' general demographic information including their age, gender, place of residence and educational background. In addition, we asked them to self-rate their sociability and WhatsApp usage on a five-point Likert scale (Low to High), and to answer four Boolean questions dealing with whether they use WhatsApp for communication with work, family, friends or others.  An IRB was obtained for ethical approval prior to beginning data collection.

We found it challenging to recruit participants, as people were quite reluctant to provide information about their WhatsApp messages, even when we emphasized that all content sent was encrypted, and that no non-encrypted content data was ever sent.  While we attempted to recruit participants from all age groups, we found that student participants, found through advertisements on campus, were the demographic most willing to participate. Nonetheless we did make a concerted effort to find people in other demographics through word-of-mouth. Through this process we recruited a total of 137 participants.  As only 19 of these participants were not college age students (18 through 34), we removed these participants' data from the analysis as this group was not large enough to be validly divided into further age subgroups. Thus, we are aware that the data collection process was biased for younger people, and hope to address this in the future through a different collection process for other age groups.

The 118 college age students were equally split between 59 men and women. However, in order to remove analysis biases from people who had not used WhatsApp for long periods of time or do not generally engage in WhatsApp conversations, we further removed another 7 people who were active in WhatsApp for under 20 days or had fewer than 10 total WhatsApp groups.   Thus, the dataset in this study contains messages from 111 participants, of which 59 were female and 52 were male, all being young adults between 18 and 34 years of age with a median of 27. The 111 participants sent and received a total of 6,449,631 messages over an average period of approximately 15 months.\footnote{The software we used collected all the data on the phone, hence the time period over which data was collected varied according to when users started using WhatsApp and their habit of deleting old messages (if at all).}

%\begin{figure}[tbp] \centering \textfloatsep2pt \intextsep5pt
%	\includegraphics[width=.7\linewidth]{Age.png}
%	\caption {The age distribution of participants in the study}
%	\label{fig:AgeDist}
%\end{figure}

The defining characteristic for the logged data is that it intentionally contains no textual content. All types of textual content are unavailable, including any special characters or emojis which exist in the messages. Similarly, we stress that we have no information about the message recipients other than an anonymous id, as all data is anonymous.

While we did not have the messages' content or recipient information, we were nonetheless still able to glean a great deal of usage information regarding message and group statistics. The first type of information focused on general information surrounding the messages' characteristics such as when they were sent, the number of words in the message, if the message included a file and the length of time that elapsed before a response was sent. Once we had all of the messages, we discretized their time into categories based on the percentages of messages sent over each hour-long interval-- e.g. messages sent between 12 and 1 A.M. Similarly, we discretized the number of messages into the categories of 1, 2, 3--5, 6--10, 11--20, and 20+ words. We then discretized messages according to the time that elapsed between messages -- under 1, 1--2, 3--5, 6--15, 16--30, and 31--60 minutes. The motivation behind this is our assumption that messages that appear within a relatively short time interval in the same group may be related to the same conversation. We emphasize that by no means does this imply that a message that appeared more than an hour after the last message was sent in a given group is not related to former messages, except that with no other supporting data (e.g., the content itself) it is impossible to make a concrete connection to prior messages. Hence, the time elapsed is the only possible, though not a perfect, indication for relevance. %After 60 minutes elapsed, we assumed the message was not a response.*** I have many reservations concerning this last topic (as appear in my notes), though I really don't know how to attack this at this point in time, other than the changes I made trying to ease the pain ***
We also discretized messages according to their file attachments and created Boolean categories of messages with and without files.

The second type of logged information concerned WhatsApp conversation groups. This dataset contained a total of 10,730 such groups from the 111 users. Note that groups with two participants are similar to a typical SMS conversation, and thus through logging this data we could test the degree to which WhatsApp has replaced traditional SMS messaging.  However, groups might also be formed around a general topic, such as a discussion about work, leisure or family issues with many more than two participants.  We logged information about the group size of all of the messages and categorized this information into the percentage of messages in trivially small groups of 2 people, groups of 3--4 participants, and those with 5 or more participants. We also collected group statistics that subsume those within the message analysis, but refer to the percentage of messages within a group having a certain attribute -- e.g. the percentage of messages sent at a certain time, of a certain length, contain a file, etc.

%Given the processed data,  we created 3 types of datasets: information related to the users' WhatsApp messages, their groups, and their overall average usage. The processed average usage information from the  111 users and their 5,245,039  messages were the sizes of these datasets. These users were part of 7,133 groups which formed the size of this dataset. In this work we analyzed all 3 datasets and found statistically significant differences in usage for people of different genders and ages across all datasets.
\section{General Analysis Methodology}
%\subsection{Dataset Analysis and Statistics}
%%%TODO for future version: It is not interesting how many messages we have in a group in-total. Instead it should be normalized according to the time messages appear in this group, i.e., the average number of messages per day.
The general methodology assumption behind this paper is that the analysis must be data-driven. As such, we use the data to support any assumptions about the nature of the data. In contrast, previous studies typically assume some type of behavior and then construct questionnaires to prove or disprove that assumption \cite{Church2013,Pielot2014,OHara2014,Mudliar2015}. In order for the data-driven approach to be successful, significant differences must be evident across different demographic groups within the data. To confirm this assumption we checked that such differences did in fact exist and were statistically significant.

\begin{figure}[tbp] \centering \textfloatsep2pt \intextsep5pt
\vspace{-15pt}
	\includegraphics[width=.62\linewidth]{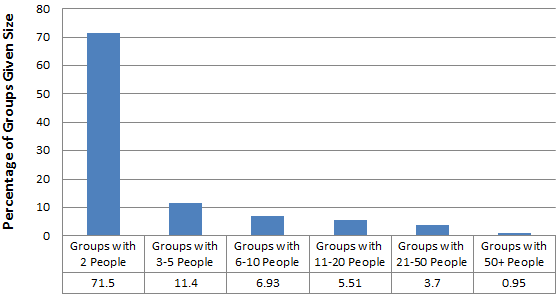}
	\caption {The distribution of the number of \textbf{groups} of each size}
	\label{fig:ConversationSize1}
	\includegraphics[width=.62\linewidth]{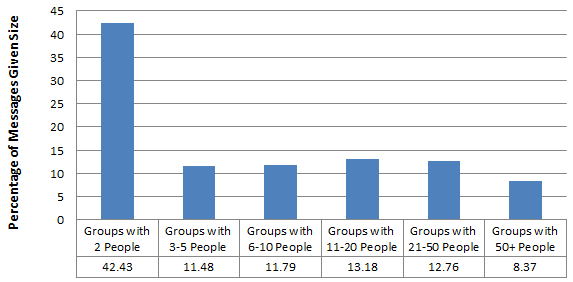}
	\caption {The distribution of the number of \textbf{messages} in each group size}
	\label{fig:ConversationSize2}
\vspace{-15pt}
\end{figure}

Specifically, we analyzed the basic distribution of messages, focusing on the statistical distributions across different genders, ages, and types of use. We found that over 70\% (71.5\%) of WhatsApp groups had only two participants (7,671 out of the 10,730), confirming previous assertions that WhatsApp is replacing SMS messaging \cite{Church2013}. On the flipside, over 50\% of all messages were not in groups of two (3,713,052 out of 6,449,631), indicating that larger groups typically were fruitful grounds for larger discussions -- something that SMS typically does not support. To better understand this point, please note these differences using the graphical distributions of the number of groups of each size in Figure \ref{fig:ConversationSize1} and the distribution of all messages in those same groups in Figure \ref{fig:ConversationSize2}.  We note that the number of groups of size two is overwhelmingly large (71.5\%), but that the number of messages in these groups is significantly smaller (42.43\%).

Please also note that while the number of groups with over 50 members is less than 1\%, these groups have a disproportionately large number of messages (8.37\%).  We believe that the reason for this is clear -- larger groups tend to have larger numbers of messages in each group. Thus, we find that a large percentage of WhatsApp activity is in fact taking the place of traditional SMS messages between two people. However, group messaging among large numbers of users, another key use of WhatsApp which SMS is less successful in supporting, also constitutes a large percentage of the WhatsApp messages we collected.

\begin{figure}[tbp] \centering \textfloatsep2pt \intextsep5pt
	\includegraphics[width=.6\linewidth]{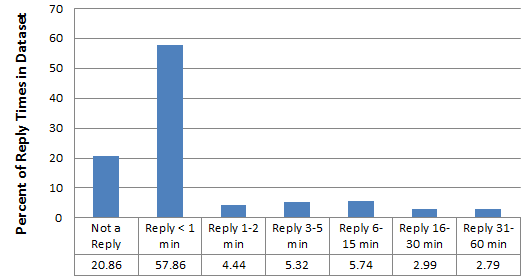}
	\caption {Analysis of reply time}
	\label{fig:ReplyTime}
	\centering \textfloatsep2pt \intextsep5pt
	\includegraphics[width=.66\linewidth]{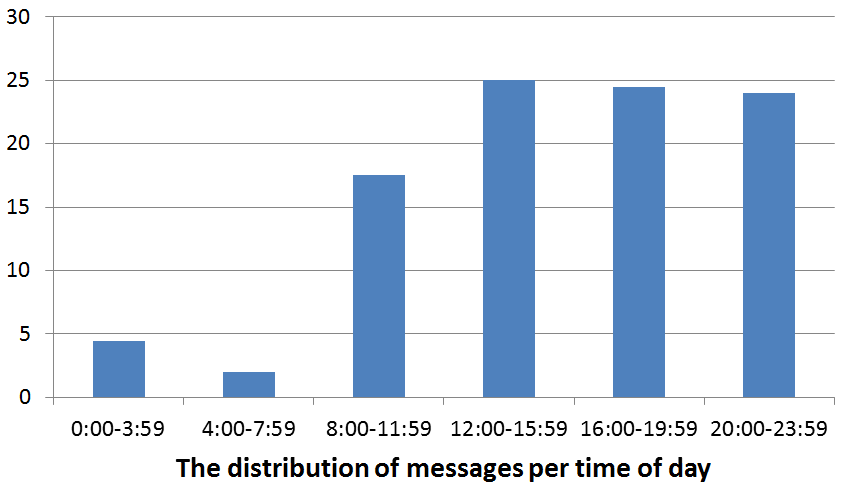} \\
	\caption {The distribution of messages per time of day}
	\label{fig:HoursStat1}
	%\vspace{-15pt}
\end{figure}

We then studied the statistical distribution of the messages' attributes starting with the average response time (time elapsed between a message and the consecutive one when in conversation), found in Figure~\ref{fig:ReplyTime}.
Please note that the average response time is quite short.
Over one half (57.82`\%) of all messages are responses that were composed within 1 minute!  This finding again confirms previous claims that WhatsApp has become a replacement to traditional SMS messaging, as most participants answer their messages quite quickly, something that is expected with SMS messaging \cite{Church2013}.

Next, we studied the distribution of the messages throughout the day (this is visually represented in Figure ~\ref{fig:HoursStat1}). As expected, very few messages were sent overnight, with under 5\% (4.36\%) being sent between midnight and 4:00 A.M. and only 2.37\% being sent between 4 and 8 A.M. Note that fewer messages were sent between 8:00 A.M. and noon (18.04\%) compared to approximately 25\% of all messages being sent in each of the other 4 hour intervals. In fact, we note no significant difference in the number of messages being sent in these three intervals (p-score $>$ 0.1), while a significantly smaller number of messages were sent between 8:00 A.M. and 12 P.M. (p-score $<<$ 0.01). Lastly, we also analyzed the message types. Here we found that most of the messages (approximately 99\%) are exclusively text messages while only 1\% included file attachments or links.

\comment
Figure~\ref{fig:MessageSize} shows the average length of all of the messages and also the average length of the participant's sent messages according to their gender.  As expected in instant mobile messaging, most of the messages are short.
About 33\% of the messages contain 1 or 2 words, another 34\% contain between 3 to 5 words, while less than 4\% of the messages contain more than 20 words.
%*** I suggested 1 word and then 2-4 and 5-10. The idea was that given what we know about the number of 1 word messages it might be way more impressive. The thing is that 1-word messages are usually less informative (stop words) and so it is a good measure for how many “non-informative” messages there are. If Sigal managed to do so (I didn't see this in the emails she sent) then we should update the figure and the text describing it, emphasizing the meaning of 1-word messages (not that the 2-word messages are not stop-messages, however the 1-word messages are very non-informative).

%\begin{figure}[tbp] \centering \textfloatsep2pt \intextsep5pt
%	\includegraphics[width=.85\linewidth]{Message.png}
%	\caption {Distribution of message length overall (above), for men (lower left) and women (lower right)}
%	\label{fig:MessageSize}
%	\vspace{-15pt}
%\end{figure}

Recall that we only possess demographic information in all datasets for the 111 known users. Nonetheless, we can isolate messages these people sent to allow us to understand different message profiles according to these people's gender and age. For example, when we analyzed the length of messages sent by men and women, we found that women send longer messages than men.
On average, women's messages include 6.5 words while men's messages include only 5.2 words.
32.86\%  men's sent messages include 1 or 2 words compared to 30.53\% for women.
At the other extreme of the distribution, 14.48\% of women sent messages that included more than 10 words while 10.11\% of the men sent messages of that length. While other works have noted that women often spend more time on WhatsApp than men \cite{Montag2015}, these results are the first to quantify these differences at the message level.

\commentend

\begin{table}[tbp] \centering \textfloatsep2pt \intextsep5pt
	%\vspace{-5pt}
	\caption{WhatsApp statistics per gender}
	\vspace{-5pt}
	\includegraphics[width=.69\linewidth]{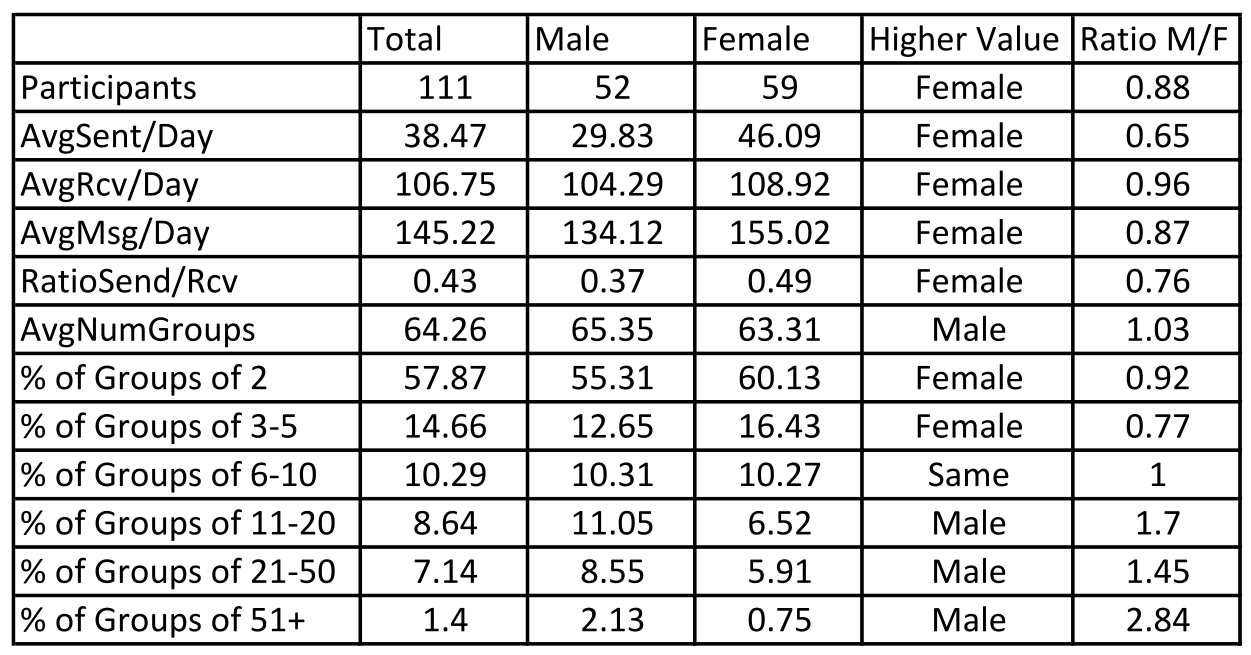}
	\label{tbl:GenderStat}
	\vspace{-10pt}
\end{table}

Table~\ref{tbl:GenderStat} contains several additional gender related insights. %*** I suggest checking statistical significance of the results that appear in the table and say something on it in the text ***
First, we found that women on average sent and received more messages than men.
Women sent and received over 155 messages a day while men sent and received approximately 134 messages (row 4), a difference of approximately 15\%.
Of these messages, women sent on average approximately 46 messages a day and received 109 messages while men sent on average slightly less than 30 messages a day and received about 104 messages (rows 2-3).  Thus, men evidently send fewer messages on average than women, something which is also evident in the differences in the ratios between sent and received messages (row 5).
Second, while on average both genders participate in a similar number of conversations (63-65 groups), the distribution of the various group sizes between the genders is different.
%*** following reported differences are very small and seem non-significant. What can you use these differences for? Are these statistically significant?  from what I saw, mostly not***
Women are more active in smaller conversation groups (60.13\% versus 55.31\% in groups with two participants), while men are more active in larger groups (11.05\% for men versus 6.52\% for women in groups of 11-20, 8.55\% versus 5.91\% in groups of 21-50 and 2.13\% versus 0.75\% in groups bigger than 50) (rows 6-12).

Overall, we found that there were significantly different WhatsApp usage patterns between different  genders and age groups. Table \ref{fig::fig3} provides details to support this claim where we present the general statistics of two different demographic groups: 1) men and women and 2) WhatsApp users younger than 25 (the median age) and aged 25 or older. We selected these age distributions based on a previous large-scale statistical analysis of WhatsApp user ages in the general population (http://www.statista.com/statistics/290447/age-distribution-of-us-whatsapp-users/). Note the differences between the average number of total messages per day (AvgMsgDay), groups (AvgGroup/Usr) per user and differences in the users' responses to the questionnaire items where they self-rated their sociality (SocialLevel), overall usage (UsageLevel), differences in the Boolean values (averaged based on values of 0 and 1), and usage in communicating with friends (UsageFriend), family (UsageFamily), and work (UsageWork). In fact, we tested all pairs of numbers for statistical significance (2-tailed t-test) and found that \textbf{all} differences were significant (p-score $<<$ 0.05) except where noted with a ``\#" at the end of each pair, as is the case of the UsageWork numbers in the pair of people 25 or older and younger than 25. Additionally, we found significant differences in the usage patterns across group usage with people who were members of these different demographics. Note the differences in the average number of minutes a user took to respond to a message (AvgResponse), the percentage of their messages which were short -- 5 words or less (Msgs5orLessWrd), the percentage of their messages which were quick responses within 5 minutes (\%RespUnder5), the average message length (AvgTextLength), and the distribution of messages across different times (midnight until 4:00 A.M., 8:00 A.M. to 12:00 P.M., and 8:00 P.M. until midnight). We also found that usage styles were different in regards to the percentage of files found in users' groups of different genders and ages (UseFile) and the percentage of groups of which they were members with 5 or more total users (isGrp5+).

We find some of the differences in Table \ref{fig::fig3} intuitive and others surprising. We are not surprised to find that younger people are more likely than older ones to send messages late at night and thus relatively older people send a higher percentage of their messages during the day. One could find support for gender differences found in people's self-rating of how much they use WhatsApp to communicate with family versus work based on previously observed differences in gender expressions \cite{Kring98}. However, we could not find a clear explanation as to why men seem to send more files in their groups than women or why older people participate in larger groups more often than younger people. These differences might point to new directions that might be confirmed with further research and questionnaires. For example, a possible hypothesis for the differences in group sizes across different ages is that younger people have more thoroughly adopted WhatsApp as a replacement for SMS messaging and consequently a larger percentage of their communication can be found in these smaller groups. % We believe that these results show differences that can spawn future research and discussion.

\begin{table*}
	\centering
	\includegraphics[width=7in]{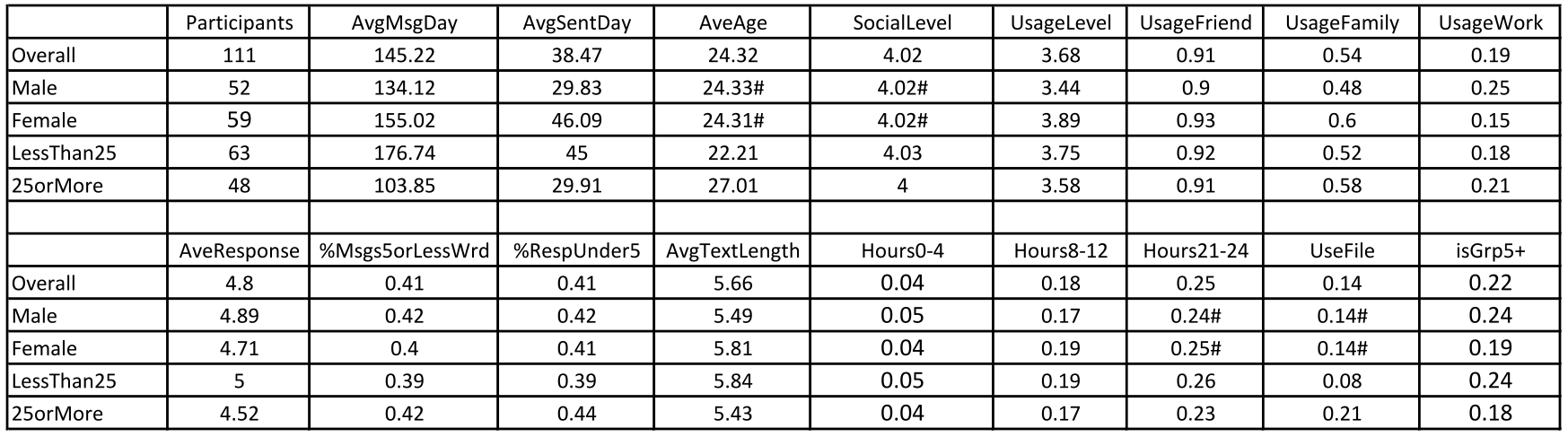}
	\caption{Results across different genders and ages in WhatsApp Dataset}
	\label{fig::fig3}
	\centering
\end{table*}

\section{Predictive Models and Hypotheses}
As we demonstrated in the previous section, significant differences do in fact exist between different types of WhatsApp users and groups. However, even statistically significant differences do not necessarily allow us to predict usage patterns. For example, the previous section demonstrated that men typically send shorter messages and women send more messages per day. However, these differences do not necessarily allow us to make a prediction about a specific user -- something that data mining algorithms do in fact allow, as we now present.

In order to illustrate the potential of using the data collected for prediction purposes, we created several predictive models for the user and group datasets, which we describe in this section.  In general, we created predictive models for \emph{user} and \emph{group} usage. User models were based on the 111 users in this dataset and were built to identify whether the author of a given set of WhatsApp posts is of a given \textbf{gender} or \textbf{age}. %Prediction models that attempt to label a person as being in these or similar categories is post of a known research direction called authorship identification. This task is traditionally defined as the process by which statistical and computational methods are used to find the author of some previously anonymous text \cite{Stamatatos2009}.

Our first hypothesis is that differences between WhatsApp authors can be predicted by exclusively using general statistics about usage, even without specific user content. In accordance with the results reported in the previous section, we posit that such differences will likely use attributes such as message length and response time as such attributes may be impacted by known gender differences \cite{Kring98}. As such, one might find that women write more in order to better express their ideas or emotions, while men write more curtly. Similarly, one might find that differences in response time or average conversation length are reflective of emotional difference -- e.g. women may prefer discussions in small groups while men prefer less personal, larger discussions. In a similar vein, one might find differences between ages, even within one gender. %*** isn't the following sentence redundant once putting this section after the demographic results section? I'm not sure ***
Such differences may be somewhat trivial, such as the time at which a message is sent -- e.g. people of certain ages might be more or less likely to work and thus be less likely to send messages at certain times - but non-trivial differences might exist too, such as differences in message length.

Our second hypothesis, based on the differences reported in accordance to the different statistics reported in the former section, is that different types of group usage can be predicted based on general group attributes, again even without considering the messages' content.  Specifically, we develop models that predict which groups will have a certain type of content such as \emph{file attachments} or \emph{shorter messages}. We also develop group models that predict which groups will have certain user activity such as a \emph{larger quantity or more frequent messages}, and \emph{quicker response times}.
%It is important to differentiate here between the group response time model and the use of message latency which is often discussed in routing literature. Other works have analyzed the time messages to be routed throughout networks, which are typically fractions of seconds \cite{Fiaa}. On the other hand, we consider the latency inherent in users' responses in the group-- something that is typically much longer than the network latency.
In theory other usage questions could have been studied, such as if a message contained certain text -- e.g. inappropriate or flagged for a certain type of content.  However, as we have no access to message content, these issues cannot be evaluated. Similarly, it may be possible that certain messages are inherently different and thus likely to be more popular or important.  Along these lines models might be created to predict which messages are apt to have certain characteristics, such as being forwarded -- something that was previously studied within the Twitter network \cite{PWebSci2011}. However, once again that study focused on the message content, which is often infeasible to rely on in real-life settings, either due to privacy or availability.

The advantage to using data mining algorithms to test these hypotheses is the objectivity of the outputted results.  On a technical level, we built models from decision trees, as implemented in the C4.5 algorithm \cite{Cs96improveduse} to create classifiers between two choices (Boolean). The C4.5 algorithm was chosen because of two main advantages. First, C4.5 identifies which attributes are most important for accurate prediction by using the InfoGain measure to rank the predictive ability of all attributes. This allows us to objectively identify which factors are most important for accurate prediction. Second, the if-then rules outputted by these algorithms allow us to observe and analyze the exact range of values within the selected attributes that form the prediction model. Furthermore, we consider many tasks, such as if a user is male / female or above / below a certain age, which are inherently Boolean decisions and are thus well suited for C4.5. In order to handle continuous attributes, we transformed the target variables into two categories through binning according to preset cutoff thresholds. For example, in creating the quick response time model, we chose a response threshold of 1 minute. We then created a Boolean classifier and assumed that anyone who answered within 1 minute answered quickly and those who answered after 1 minute, even if they answered only seconds after 1 minute, did not. More specifics of the models and their findings are in the next section.

\begin{table*}
	\includegraphics[width=7in]{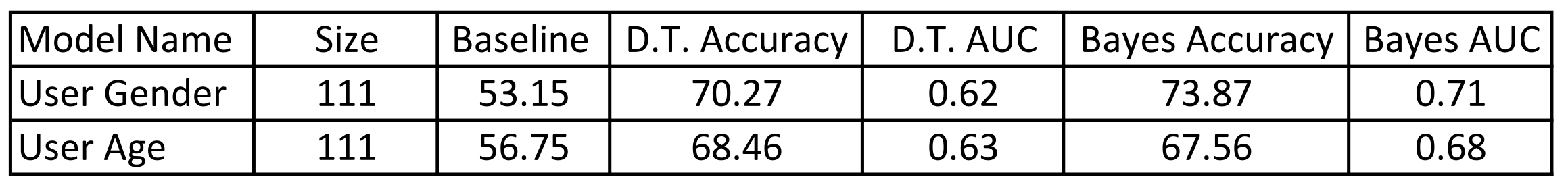}
	\caption{Authorship identification prediction of Gender and Age based on average WhatsApp user data}
	\label{fig::fig4}
	\centering
	\includegraphics[width=7in]{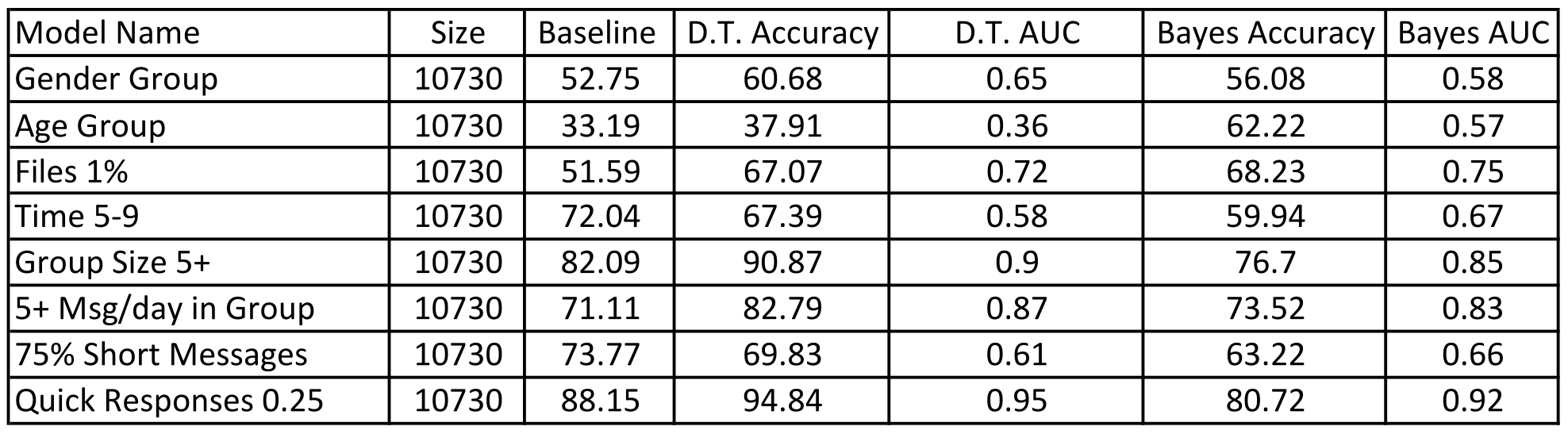}
	\caption{Predicting group activity in WhatsApp Dataset}
	\label{fig::fig5}
\end{table*}

\section{Data Analytic Results}

In general, we built two types of models using the popular open source Weka data mining package \cite{Weka} -- decision trees and probabilistic models based on Bayesian networks. We did consider other models, but as the Bayesian models often did better than the other alternative algorithm, we present results from this algorithm for comparison. Within the user models, standard 10-fold cross validation was used to assess all models as the validation set was always a different set of users than those used in the training data. While we also considered using standard cross-validation in order to assess the group prediction models, we rejected this approach as at times we noted that both the training and testing datasets contained groups from the same user. Instead, for each group model we generated 10 randomized splits which ensured that a user's groups were only within the training or the testing dataset. While the resultant stratified training-testing splits were not always of the same size, they did guarantee we did not overfit by having the same user in both the training and testing data.

%  71.5267 \ second age model

Overall, we were successful in predicting an author's gender and approximate age based on users' general data, as can be seen in Table \ref{fig::fig4}. %We present results from decision tree and Bayesian network models from both the user and group datasets.
The first column in the table presents how many records were in each dataset. The next two columns present the accuracy and Area under the Curve (AUC) of the decision tree model with the following two columns presenting the accuracy and AUC of the corresponding Bayesian model. The first row presents the results for predicting gender based on the data and the second row presents the results for predicting age -- e.g. 25 or older versus under 25 years old. This cut-off was chosen as it represented roughly a 50-50 split within the data. Minimally, a successful model should at least be more accurate than this value. Note that both models were successful in both tasks as the predictions' accuracies were much greater than the baseline values.

\begin{figure}[tbp] \centering \textfloatsep2pt \intextsep5pt
\includegraphics[width=.6\linewidth]{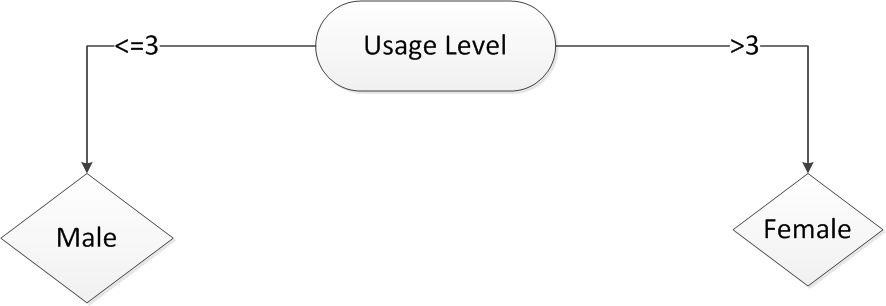}
\caption {Predicting Male / Female from all collected data}
\label{fig:tree1}
\includegraphics[width=.6\linewidth]{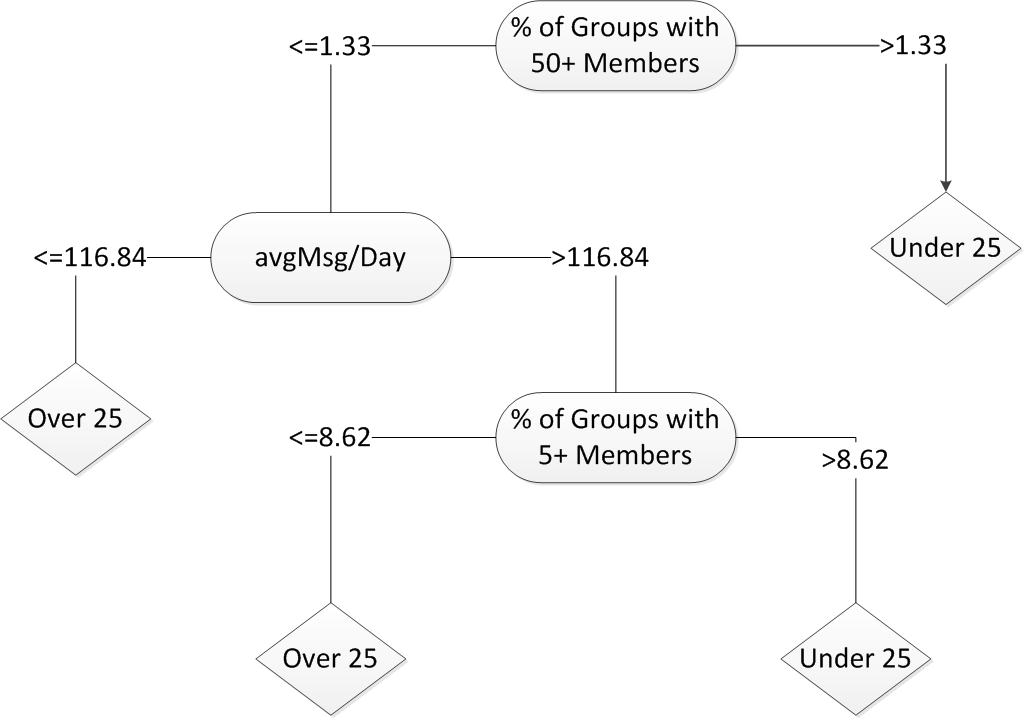}
\caption {Predicting author age from user data}
\label{fig:tree2}
\vspace{-5pt}
\end{figure}
 An advantage in building decision trees is noting the logical rules and the attributes in the learned models. We studied this output from the decision tree models. We noted that the decision tree for predicting gender focuses on the usage level, the response time and the number of large groups a user had. Specifically, we found that men overall self-reported lower usage levels, on average took longer to respond, and have large groups. A slightly simplified version of the gender decision tree is presented in Figure \ref{fig:tree1}. Note that this rule is relatively simple: if the user self-rated a usage level of 3 or less they were male, otherwise they were female. Despite the simplicity of this decision tree, it still yielded an accuracy of 63.56\%. Nonetheless, more complex models could be built with both decision trees and Bayesian networks as reported in Table \ref{fig::fig4}.  %Referencing Table \ref{fig::fig3} note that while several differences in the values are evident, the difference in average UsageLevel was most significant and thus chosen as the root of the decision trees.

 Using a similar methodology, we were able to differentiate between users below the age of 25 and those above it. Here, we noted that younger people had more messages a day (AvgMsgDay) and were likely to be in groups with more than 5 people (isGrp5+).  The decision tree for predicting age, found in Figure \ref{fig:tree2}, shows the exact rules behind this classifier. The model predicted that if a person had more than 1.33\% of all groups belonging to a group of 50 or more members they were under 25, but if they had fewer than this number of large groups and received on average less than 117 messages per day they were 25 or older. Otherwise, a third rule was needed to differentiate between younger people, with more than 8.62\% of their groups constituting 5 or more people, and older people, with fewer groups of this size. We again note that while these rules are consistent with the general trends seen in Figure \ref{fig::fig3}, the decision tree provides a predictive model with exact thresholds that predict differences between the groups.

We also created models for the gender and age prediction tasks using the group dataset, the results of which are in the first two rows of Table \ref{fig::fig5}. As we overall have much more group data (10,730 records) than average user data (111 users), it is not surprising that these models performed better than the user models, particularly when noting the differences in AUC. Again, the decision trees provide insight as to which attributes are most helpful. In the first decision tree we found that once again men were characterized by lower usage levels, use WhatsApp less for family communication, and have shorter messages. Within the age classification task we found that younger people use files less frequently than older people and are less likely to use WhatsApp for family and work communication. While the attribute AvgMsgSent played prominently in the classification task from the user dataset, this attribute was absent from the group dataset as average statistics for a user are not evident from a group's profile. Nonetheless the group statistics proved to be even more helpful in building age and gender models.

We also built models that predicted group usage characteristics, the results of which are also found in Table  \ref{fig::fig5}. Specifically, we built models to predict which groups will contain file attachments in at least 1\% of all messages (row 3), will have more than 25\% of their messages sent between 5:00 and 9:00 A.M. (row 4), which messages are characteristic of groups with 5 or more users (row 5), will average at least a total of 5 messages send a received that per day (row 6), will on average contain short texts with five words or less in at least 75\% of all messages (row 7), and will have at least one quarter of all messages responded to within one minute (row 8). Please note that in general these models were much more successful than the Baseline values, with AUC values often above 0.8. However, some exceptions do exist. Note that predicting age based on group activity was not successful within the decision tree model with an AUC value of below 0.5 (0.36). Nonetheless, even here the Bayesian model was more successful, with an accuracy of nearly 30\% greater than the baseline and an AUC above 0.5. The thresholds used in this task were meant to be representative of which groups are more active in the morning, have smaller messages, shorter response times, etc. We did in fact check other tasks (e.g. messages sent later at night, in larger groups, with difference thresholds for the group sizes and response times, etc.) and found that the data similarly supported prediction beyond the specific thresholds reported in this paper.

\begin{figure}[tbp] \centering \textfloatsep2pt \intextsep5pt
\includegraphics[width=.65\linewidth]{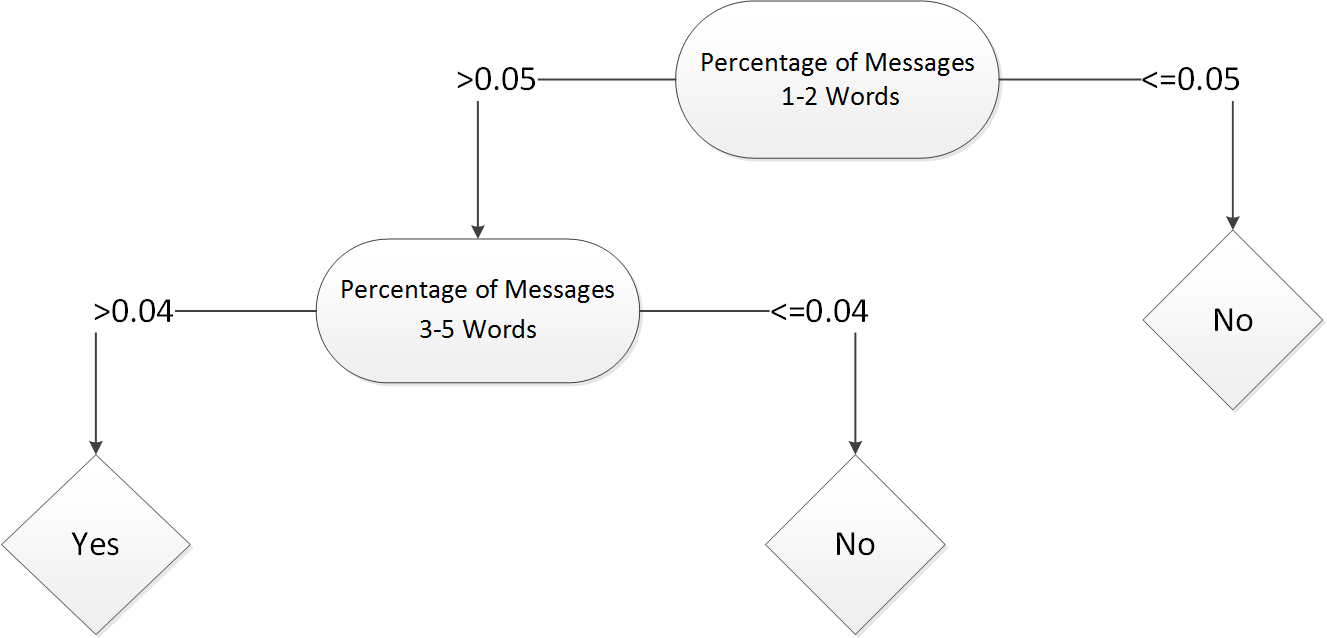}
\caption {Predicting if one quarter of the messages are answered within one minute }
\label{fig:tree3}
\vspace{-5pt}
\end{figure}
% 71.5267 \%

The decision tree models also facilitated the ability to understand which attributes were most influential in predicting the group's behavior. For example, a simplified version of the decision tree to predict which groups had one quarter of the messages answered within a minute is found in Figure \ref{fig:tree3}.  Specifically, groups with a larger percentage of shorter messages (1-2 words and 3-5) typically indicate quick answers. Similarly we were able to use decision trees to understand the models for other group behaviors.  We found that groups with more files were composed of young participants with advanced schooling (M.S. or more), yet are 28 or younger. Additionally, groups with users who didn't rate themselves with high usage levels, but had high educational levels (above 16 years) and were above 30 still typically sent more file attachments. As one might expect, we also found that full time students were less likely to be active in the morning in comparison to those who had jobs.  As Figure \ref{fig:ConversationSize2} demonstrates, we also found that groups with 5 or more participants have more messages and thus typically have messages sent with a higher frequency. We found that larger groups typically contained shorter messages. We also found that younger people typically send shorter messages and while older people typically send longer messages, they do so less frequently.

\section{Conclusions and Future Work}
This work represents the first exhaustive analysis of WhatsApp messages.  We collected over 5 million messages from over 100 students between the ages of 18 to 34, and differentiated between different types of \emph{user} and \emph{group} usage of the network.  A key characteristic of this study is that we did not collect or analyze any content within the messages. This was done intentionally to safeguard participants' privacy. Despite this limitation, we found that many message and group characteristics significantly differed across users of different demographics, such as gender and age, and present these results through performing extensive statistical analysis. Additionally, we believe that one key novelty of this work is that we use data analytics to predict users' gender, age and group activity. As our work is data driven, we base our findings on the algorithms' output, and did not attempt to verify any specific thesis as had been previously done.  This is one key advantage to using data analytics, and this difference is especially clear from the decision tree results presented in this paper.

Overall, our results provide several new insights into WhatsApp usage.  We find that the younger users in this dataset used this network more frequently.  We also find that more years of education and age are positive factors in predicting how frequently people send file attachments. Overall, women use this network more often than men and they reported that they use it more often to both generally communicate and communicate with family. Men, on the other hand, are generally members of larger communication groups and send shorter messages. Additionally, larger groups are not only defined by their large number of users, or even the large numbers of messages that are frequently sent, but are also typically defined as having shorter messages than those in private one-to-one communications. Decision tree models were not only helpful in identifying these attributes, but were useful in providing the thresholds within the if-then rules for the models that predicted these results. As our results are built through analyzing users' general message data, but without their content, we believe the methodology used in our analysis may be of general interest to other groups such as demographers and government bodies to facilitate data analysis without infringing on users' privacy.

In building upon this work, we believe that two types of studies will likely lead to fruitful results. First, we believe that additional studies should be undertaken to improve upon and extend the study we present. While this study analyzed over 5 million messages, it is still somewhat limited in containing only 111 users and exclusively focusing on people between the ages of 18 and 34. Furthermore, we believe it will be helpful to study how different demographic groups use WhatsApp. We believe that even more accurate models can be built through studying data from more users, with a wider range of ages and different ethnic backgrounds. Similarly, we did not study all group tasks, and other tasks - such as which messages will be forwarded - remain unexplored. In a related matter, while we intentionally built models without analyzing user content in order to safeguard privacy, even more accurate models might be built in the future if user consent could be obtained for this information.

We believe a second type of direction should focus on applying the lessons learned from this paper's models. It may be wise to customize user interfaces for certain types of users and tasks based on the attributes found to be important in this paper. For example, users who are more educated or older might prefer a different WhatsApp interface compared to less educated or younger users as their usage patterns differ significantly. Similarly, as larger groups are characterized by shorter messages, it may be that the interface for these types of interactions should be customized with this information in mind as well. We hope that these and other issues will be explored in greater detail in future work.

\bibliographystyle{alpha}
\bibliography{biblio}
\end{document}